\documentclass[a4paper,10pt,twoside]{cpc-hepnp}

\usepackage{multicol}
\usepackage{graphicx}
\usepackage{booktabs}
\usepackage{amssymb,bm,mathrsfs,bbm,amscd}
\usepackage[tbtags]{amsmath}
\usepackage{lastpage}
\usepackage{multirow}
\usepackage{booktabs}
\usepackage{threeparttable}
\newcommand{\tabincell}[2]{\begin{tabular}{@{}#1@{}}#2\end{tabular}}

\begin{document}

\fancyhead[c]{\small Submitted to ¡®Chinese Physics C¡¯      Chinese Physics C~~~Vol. 37, No. 1 (2013)
0346} \fancyfoot[C]{\small 010201-\thepage}

\footnotetext[0]{Received 9 December 2013}

\title{Monte Carlo Simulation of Indoor External Exposure due to Gamma-emitting Radionuclides in Building Materials}

\author{%
      DENG Jun$^{1)}$\email{djoy0118@gmail.com}%
\quad CAO Lei$^{}$
\quad SU Xu$^{2)}$\email{suxu@nirp.cn}%
}
\maketitle

\address{%
Key Laboratory of Radiological Protection and Nuclear Emergency, Chinese Center for Disease Control and Prevention,\\
National Institute for Radiological Protection, China CDC, Beijing 100088, China\\
}

\begin{abstract}
The use of building materials containing naturally occurring radionuclides£¬such as $^{40}$K, $^{238}$U , $^{232}$Th and their progeny, could lead to external exposures to the residents of such buildings. In this paper, a set of models are constructed to calculate the specific effective dose rates(the effective dose rate per Bq/kg of $^{40}$K, $^{238}$U series, and $^{232}$Th series) imposed to residents by building materials with MCNPX code. Effect of chemical composition, position concerned in the room and thickness as well as density of material is analyzed. In order to facilitate more precise assessment of indoor external dose due to gamma-emitting radionuclides in building materials, three regressive expressions are proposed and validated by measured data to calculate specific effective rates for $^{40}$K, $^{238}$U series and $^{232}$Th series, respectively.
\end{abstract}



\footnotetext[0]{\hspace*{-3mm}\raisebox{0.3ex}{$\scriptstyle\copyright$}2013
Chinese Physical Society and the Institute of High Energy Physics
of the Chinese Academy of Sciences and the Institute
of Modern Physics of the Chinese Academy of Sciences and IOP Publishing Ltd}%

\begin{multicols}{2}

\section{Introduction}

Radioactive nuclides have always been present in the natural environment, which make the major contribution to the annual average doses of radiation exposure of the world population. Building materials containing $^{40}$K, $^{238}$U series and $^{232}$Th series, play a significant role in indoor external dose besides terrestrial and cosmic radiation, as 80\% of the human life time is spent at home and/or office. Indoor elevated external dose rates may arise from and relate to high activities of radionuclides in building materials\cite{lab1}. Hence, it is not only important but also feasible to assess the radiological hazard by calculating indoor external dose based on radioactivity measured for building materials.

Since the first model to calculate external dose rate due to building materials was proposed by Koblinger in 1978\cite{lab2}, many alternative ones have been published either based on Monte Carlo or on analytical method~ \cite{lab3}-\cite{lab10}. Among them, the quantity of specific absorbed dose rate in air was defined as the absorbed dose rate in air due to mass activity of 1 Bq/kg of the parent and any progeny in the building materials. Although some outliers were met in the literature, ranging from £­20\% to +30\%, the values seem to converge to 0.080, 0.89, and 1.02 nGy/h per Bq/kg for $^{40}$K, $^{238}$U series and $^{232}$Th series, respectively.

However, the values above are constrained to the standard room consisting of concrete wall with the thickness of 20 cm and density of 2.35 g/cm$^{3}$. The specific absorbed dose rate in air is significantly influenced by several parameters as position concerned in room, thickness and density of building materials etc., which has been demonstrated in several studies~\cite{lab1,lab11}. In the present study, the effect of position concerned in room, thickness and density as well as chemical composition is studied, and three regressive functions to calculate the specific effective dose rate induced by building materials are obtained based on MCNPX code for $^{40}$K, $^{238}$U series and $^{232}$Th series, respectively.

\section{Materials and methods}

\subsection{Room geometry}
According to typical living premises, a room with the inner dimensions of 4 m$\times$5 m$\times$2.8 m has been assumed. Walls, ceiling and floor have the same thickness ranging from 5 to 35 cm with the increase step of 5 cm. The room has one door (2 m$\times$0.9 m) and one window (1.5 m$\times$1.5 m) on the opposite wall. A schematic drawing of the room and the detector position is shown in Fig.~\ref{fig1}. Various commonly used building materials, such as ordinary concrete, autoclaved aerated concrete, clay brick and marble as well as granite that filled in the wall, ceiling and floor, are studied.

\begin{center}
\includegraphics[width=7cm]{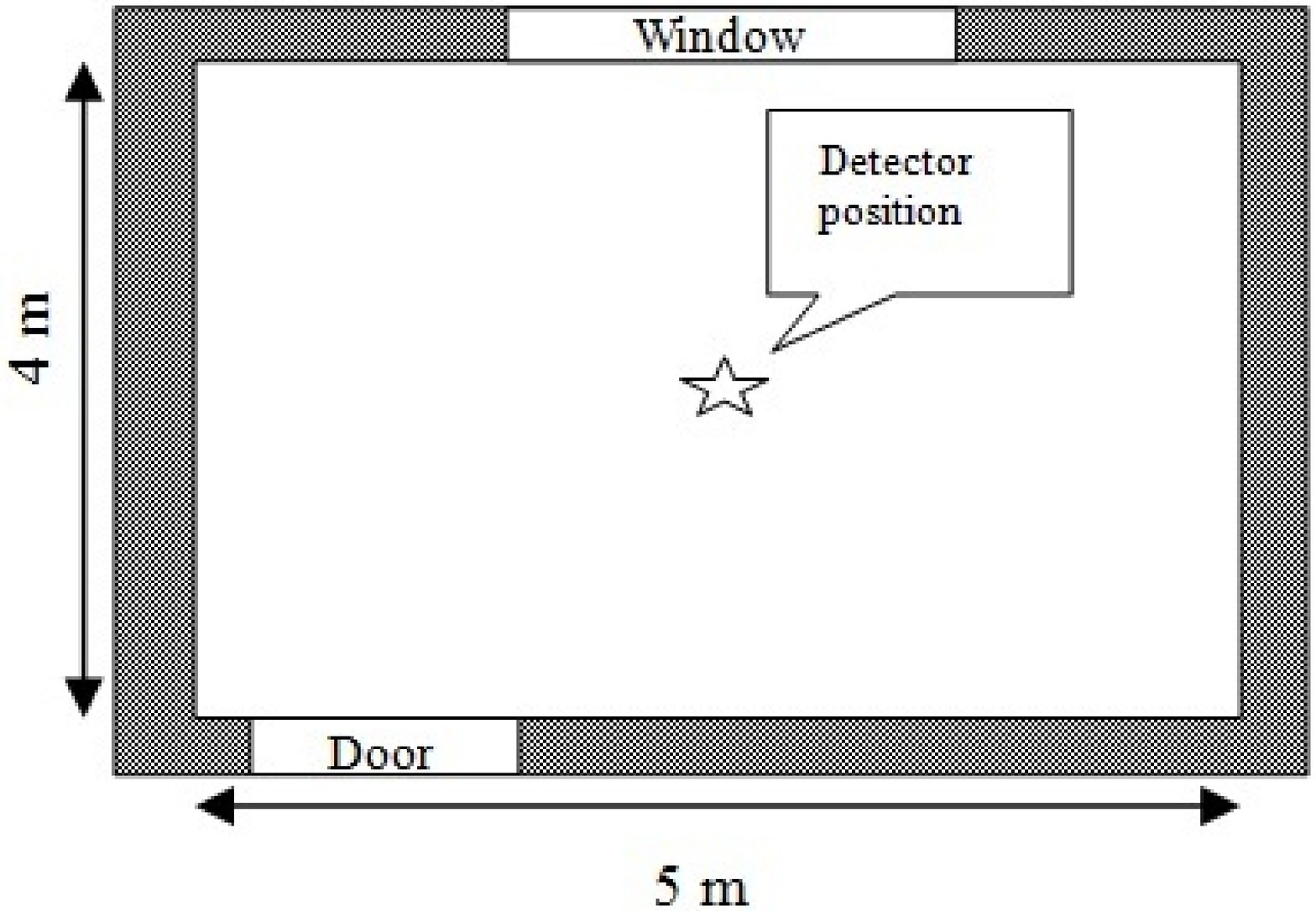}
\figcaption{\label{fig1}  Ground view of the room assumed to simulate indoor external exposure.}
\end{center}

\subsection{Monte Carlo simulation}
Photon transport in walls and in air inside the room is simulated by MCNPX code, which is an ideal platform for modeling of the passage of particles through matter. The photon flux and energy deposition in the target air regions are scored using track-length estimator\cite{lab12}.

The modeled radioactive sources are assumed to be uniformly distributed in walls, ceiling and floor, and to emit isotropic photons only. Both $^{238}$U series and $^{232}$Th series are assumed to be in secular equilibrium. To ensure the calculation efficiency, only the photons with discrete energies greater than 10 keV and emission intensity greater than 1\% of $^{238}$U series and $^{232}$Th series decay chains are accepted as the input data for simulation model, as shown in Fig.~\ref{fig2}. $^{40}$K is set to emit photon with the energy of 1.46 MeV and intensity of 10.7\%. The energy and yields of the sources are taken from electronic version of the ICRP Publication 38\cite{lab13}.

\begin{center}
\includegraphics[width=7cm]{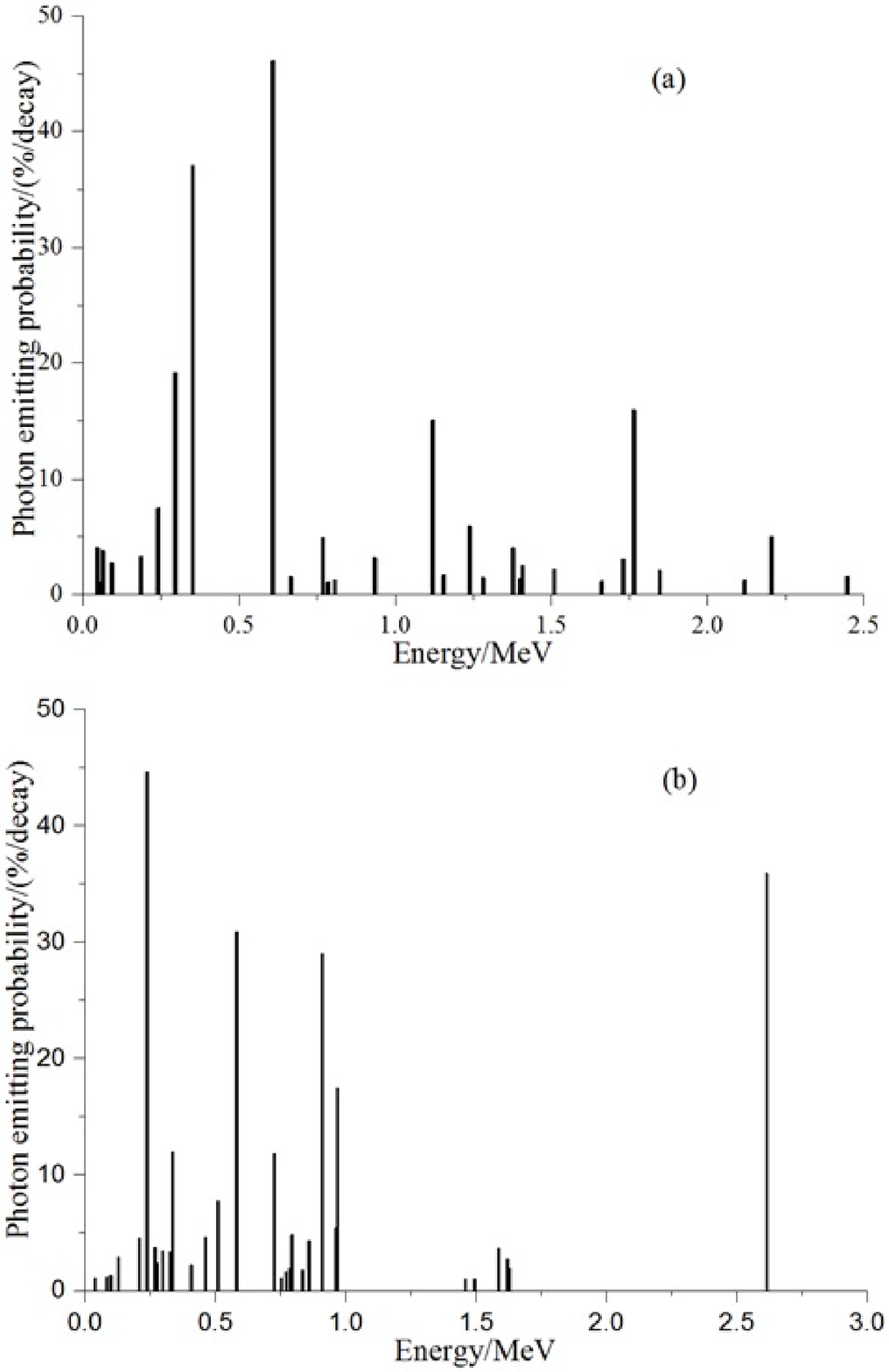}
\figcaption{\label{fig2}  Input photon spectra in simulation model for (a) $^{238}$U series;(b) $^{232}$Th series.}
\end{center}

The point-detector, F5 tally, is set up to score the flux at the point in the middle of room at the height of 100 cm. As the endpoints of the simulation are effective dose rates per unit activity concentration of the source nuclides in building materials, the DE/DF function is employed to convert flux to effective dose. The effective dose conversion coefficients per fluence for monoenergetic photon sources in several simplified irradiation geometries are defined in ICRP Publication 116\cite{lab14}. As isotropic geometry is most relevant to the considered irradiation conditions inside the room, thus coefficients for isotropic geometry have been used in the simulation. The simulation particle number is 3$\times$10$^{7}$ and all the relative errors of simulated results are less than 2\%.

\section{Results and discussion}

\subsection{Validation of the model established}
Values of absorbed dose rate in air per unit mass concentration of the source radionuclides obtained in this work and parameters adopted in simulation as well as those found in other similar literatures are compared in Table~\ref{tab1}.

In Table~\ref{tab1}, results are included that were obtained by Koblinger, who used multi-group adjoint Monte Carlo simulation to calculate specific dose rates in air at a point inside a compartment with dimensions of 4 m$\times$5 m$\times$2.8 m, defined by three pairs of homogenous 20 cm-thick concrete walls\cite{lab3}. This setup was considered as a standard for comparison by the authors who followed Koblinger. For example, several authors considered the same compartment but applied various analytical techniques to compute dose rates\cite{lab4}-\cite{lab10}. Their results are also included in Table~\ref{tab1} for comparison with results of the present study.

As seen from Table~\ref{tab1}, the results for $^{40}$K and $^{232}$Th series differ insignificantly, within -6\%-3\%, with the average values above reported by quoted authors for the standard room. The value for $^{238}$U series shows the difference of -14\%, with the counterpart one. Besides obviously different computational methods and input data used, these difference can be also attributed to presence of the door and the window, and the differences in the input energy spectra. For example, the room model set up by Koblinger filled with pure SiO$_2$ with a density of 2.32 g/cm$^{3}$, without doors and windows. In views of the variation coefficients of 13\%-15\% in the reported specific absorbed dose rates, the value obtained in present work can be considered as acceptable, namely the simulation model established is reliable.

\end{multicols}

\begin{center}
\tabcaption{ \label{tab1}  Comparison of specific absorbed dose rates in air
according to various publications and calculation in the present work.}
\footnotesize
\begin{tabular*}{170mm}{@{\extracolsep{\fill}}cccccccc}\toprule
\multirow {2} {*} {References}&\multirow {2} {*}{\tabincell{c}{Method\\used}}&\multirow {2} {*}{\tabincell{c}{Room \\dimension/m$^{3}$}} &\multirow {2} {*} {\tabincell{c}{Wall \\thickness/m}}&\multirow {2} {*} {\tabincell{c}{Density/\\ £¨kg/m$^{3}$£©}}&\multicolumn {3}{c}{\tabincell{c}{Specific absorbed dose rates in air/\\((nGy/h)/(Bq/kg))}} \\
\cline{6-8}
                         &  &  &  &  & $^{40}$K&$^{238}$U series&$^{232}$Th series\\
\hline
Koblinger (1978)         &Monte Carlo&4¡Á5¡Á2.8&0.2&2320& 0.078 & 0.93 & 1.03  \\
Stranden (1979)          &Analytical &4¡Á5¡Á2.8&0.2&2350& 0.078 & 0.92 & 1.11  \\
Mustonen (1984)          &Analytical &4¡Á5¡Á2.8&0.2&2350& 0.078 & 0.89 & 1.05  \\
Markkanen (1995)         &Analytical &4¡Á5¡Á2.8&0.2&2350& 0.077 & 1.06 & 0.91  \\
Ahmad et al.(1998)       &Analytical &4¡Á5¡Á2.8&0.2&2350& 0.081 & 0.95 & 1.21  \\
Maduar and Hiromoto(2004)&Analytical &4¡Á5¡Á2.8&0.2&2350& 0.072 & 0.70 & 0.92  \\
Ademola and Farai (2005) &Analytical &4¡Á5¡Á2.8&0.2&2350& 0.080 & 0.88 & 1.19  \\
Al-Jundi et al.(2009)    &Monte Carlo&4¡Á5¡Á3  &0.2&2350& 0.067 & 0.69 & 0.91  \\
Moharram et al.(2012)    &Monte Carlo&4¡Á5¡Á2.8&0.2&2350& 0.111 & 1.02 & 0.87  \\
Average of values above  &           &       &   &    & 0.080 & 0.89 & 1.02  \\
Variation coefficient    &           &       &   &    & 15\%  & 14\% & 13\%  \\
This work                &Monte Carlo&4¡Á5¡Á2.8&0.2&2350& 0.076 & 0.782& 1.058 \\ \bottomrule
\end{tabular*}%
\end{center}

\begin{multicols}{2}
\subsection{Calculation of specific effective dose rate}
The specific effective dose rates for $^{40}$K, $^{238}$U series and $^{232}$Th series which assumed uniformly distributed in ordinary concrete, autoclaved aerated concrete (AAC), clay brick, marble and granite are obtained and compared in Table~\ref{tab2}. The results show that the chemical compositions of the building materials have no significant effect on the specific effective dose rate, on condition that the densities and thickness remain the same. This is in accordance with Kobliger who also found no significant differences between various building materials with the same surface density\cite{lab2}. Therefore, the specific effective dose rate determined for concrete are also valid for other building materials.

\end{multicols}

\begin{center}
\tabcaption{ \label{tab2}  Specific effective dose rates obtained for various building materials.}
\footnotesize
\begin{tabular*}{170mm}{@{\extracolsep{\fill}}ccccccc}\toprule
\multirow {2} {*} {Building materials}&\multirow {2} {*}{\tabincell{c}{Wall \\thickness/£¨m£©}}&\multirow {2} {*}{\tabincell{c}{Density/\\(kg/m$^{3}$)}} &\multirow {2}{*}{\tabincell{c}{Surface density/ \\£¨kg/m$^{3}$£©}}&\multicolumn {3} {c} {\tabincell{c}{Specific effective dose rates/\\((nSv/h)/(Bq/kg))}} \\
\cline{5-7}
 & &  &  & $^{40}$K & $^{238}$U series&$^{232}$Th series\\
\hline
Ordinary concrete      &0.2&2350&470& 0.054 & 0.522 & 0.745  \\
Clay brick      &0.2&2350&470& 0.053 & 0.523 & 0.746  \\
AAC$^{1}$            &0.2&2350&470& 0.053 & 0.523 & 0.744  \\
AAC$^{1}$            &0.2&500& 100& 0.021 & 0.217 & 0.91  \\
AAC$^{2}$            &0.2&500& 100& 0.021 & 0.218 & 0.300  \\
AAC$^{3}$            &0.2&500& 100& 0.021 & 0.217 & 0.299 \\
Marble          &0.02&2350& 47&0.012 & 0.125 & 0.173 \\
Granite         &0.02&2350& 47&0.012 & 0.125 & 0.173  \\ \bottomrule
\end{tabular*}
\end{center}

\begin{tablenotes}
\item[1] 1 AAC consists of cement, fly-ash and lime;2 AAC consists of cement, slag and sand;3 AAC consists of cement, sand and lime.
\end{tablenotes}

\begin{multicols}{2}

The position at the standard height of 100 cm is usually chosen as a typical reference point for the dose to an inhabitant staying in a room. To verify whether the position is of major importance, the values for various positions at 100 cm above floor are assessed by using tmesh tally. For that purpose, the inner room space at the height from 95 to 105 cm is divided into 80 equal cells measuring 50 cm$\times$50 cm$\times$10 cm; the averaged specific effective dose rates for each cell are obtained and shown in Fig.~\ref{fig3}. As shown in the Fig.~\ref{fig3}, the specific effective dose rates of 80 cells differ insignificantly except those values decline following the position approaching the door and window. Those values for 80 cells are put together resulting in the average and standard deviation of 0.053¡À0.004 (variation coefficient: 7.5\%), 0.524¡À0.027 (variation coefficient: 5.2\%), 0.750¡À0.038 (variation coefficient: 5.1\%), for $^{40}$K, $^{238}$U series and $^{232}$Th series, respectively, which almost equal to those in center position at the same height. The relative percentage deviation between values obtained in the middle of the room and the average ones are 1.8\%, -0.4\% and 0.7\%, which are all within the variation coefficient of 80 values. In addition, the position of habitant in the room can be anywhere inside the room, entirely random. Therefore, it is reasonable to ignore the effect of position while assessing inner external dose due to building materials.

The thicker and denser a construction part is, the more radioactivity it contains, and therefore these two parameters play an important role in evaluating the value of the specific effective dose rate. The specific effective dose rates of various materials with different thickness from 2 to 35 cm and densities of 0.5, 2.35 and 2.7 g/cm$^{3}$ are calculated with the standard room geometry (inner room dimension of 4 m$\times$5 m$\times$2.8 m with window and door). Results show that the specific effective dose rates differ significantly due to the effect imposed by thickness and density, indicating that it is unreasonable to assess the indoor external dose ignoring the effect imposed by thickness and density of the building materials. This has been proposed in several publications\cite{lab12,lab16}.

\begin{center}
\includegraphics[width=7cm]{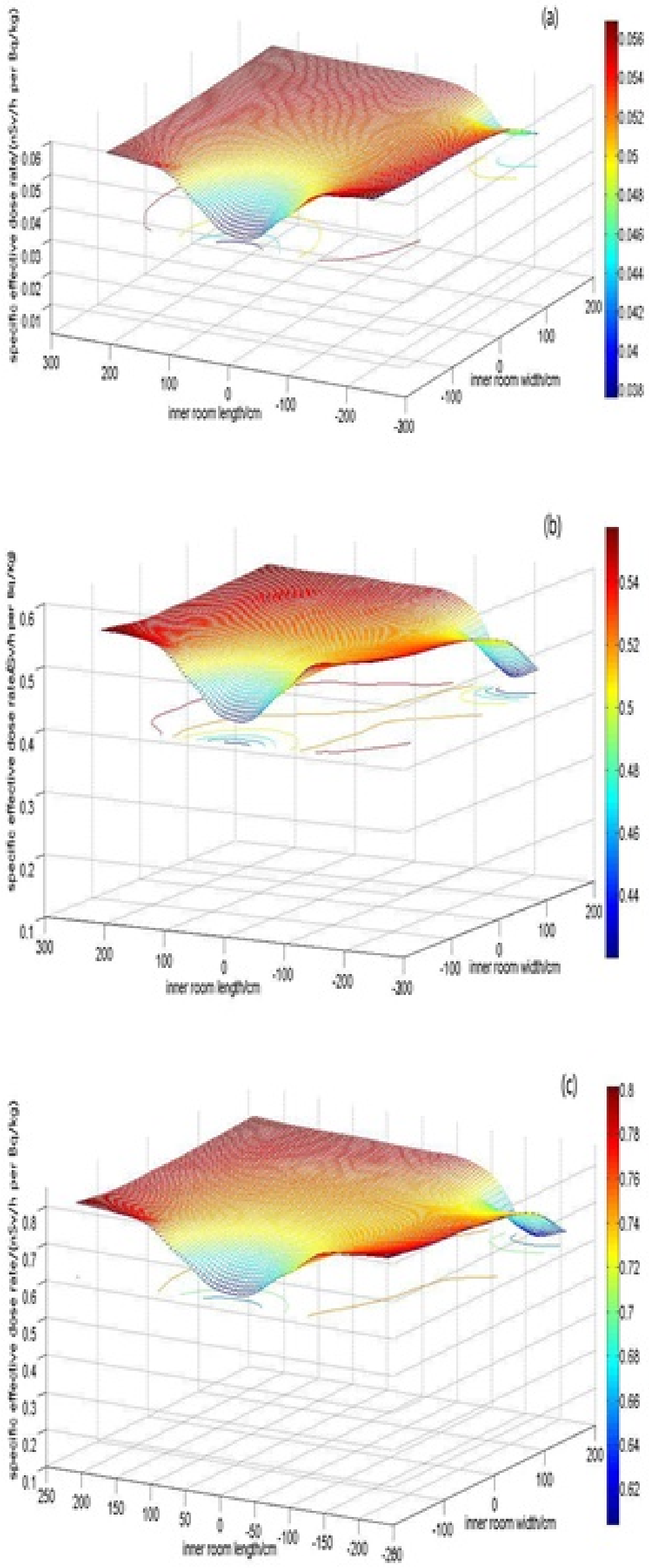}
\figcaption{\label{fig3}  Specific effective dose rates for various positions inside the room
 induced by (a) $^{40}$K;(b) $^{238}$U series;(c) $^{232}$Th series.}
\end{center}

In order to establish a material-independent approach to facilitate the inner room dose assessment, both variables are combined to the surface density in present paper. Consequently, the specific effective dose rate can be calculated by following expressions as shown in Fig.~\ref{fig4}, which are obtained by conducting a non-linear regression, for $^{40}$K, $^{238}$U series and $^{232}$Th series, respectively. As shown in Fig.~\ref{fig4}, good accordance between regression curves and the simulated data is obtained.

\begin{center}
\includegraphics[width=7cm]{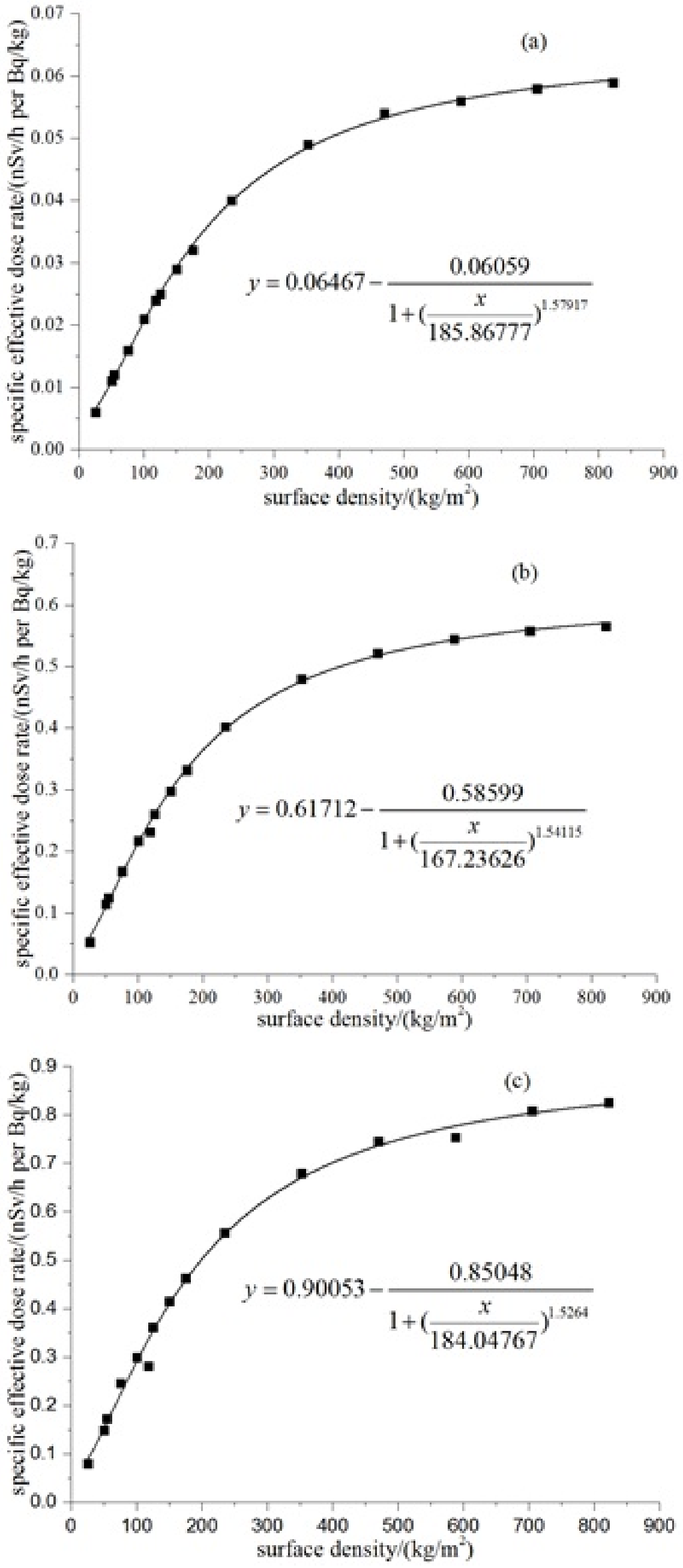}
\figcaption{\label{fig4}   Regressive curve for calculating specific effective dose rate inside the room
induced by (a) $^{40}$K;(b) $^{238}$U series;(c) $^{232}$Th series.}
\end{center}

\subsection{Comparison with the measured data}
Calculations are carried out and measurements are made for dwellings constructed from concrete. The typical size of the room selected in the buildings is 3.7$\times$4.5$\times$2.6 m$^{3}$, the wall, ceiling and floor thickness are 12 cm. The specific effective dose rates for a single room, as calculated based on regressive expressions, are 0.044, 0.436, and 0.609 nSv/h per Bq/kg.

 The specific activity of the building materials is measured by Hand-Held radioisotope identifiers with HPGe detector and Gamma Vision program for Windows. The average values of 7 similar rooms are 215.3, 23.6 and 38.9 Bq/kg for $^{40}$K, $^{238}$U and $^{232}$Th, respectively. From the calculated specific effective dose rates and the measured activities, a total irradiation dose rate of 56.1nSv/h is expected in the center of a single room, with the influence from adjacent apartment being corrected~\cite{lab2,lab15}.

To check the result, measurements are also made using portable X/¦Ã dose rate instrument. In 12 different floors of various buildings, the effective dose rates varied from 61.3 to 73.8 nSv/h, with an average of 67.7 nSv/h (the coefficients for ambient dose equivalent to effective dose are taken from ICRP 116\cite{lab14};the contribution of cosmic radiation is extracted from the measured data\cite{lab17}).

The agreement between the calculated and measured data is satisfactory if the error of the measurements, the approximations of the model and the statistical error of the calculations are taken into account. Another possible source of error, although slight, is the ¦Ã rays in the air emitted by radon-daughters which are assumed constrained inside the building materials.

\section{Conclusions}
In present work, a set of room models were set up to simulate the specific effective dose rate induced by radionuclides contained in building materials, based on Monte Carlo method. Influence on the value by several parameters such as chemical composition, position concerned of reference in the room and thickness as well as density of building materials was analyzed and weighted. Results show that other than chemical composition and position concerned in the room, density and thickness which can be combine to surface density, are of major importance to assess indoor external dose to habitant in dwelling.

Three regressive equations based on surface density to calculate the specific effective dose rates due to $^{40}$K, $^{238}$U series and $^{232}$Th series were suggested and validated by on site measurement data, for more precise assessment of indoor external dose to the inhabitant by radionuclides in building materials.

\end{multicols}

\vspace{10mm}

\begin{multicols}{2}

\end{multicols}

\clearpage

\end{document}